\documentclass[aos,nameyear,dvips]{arximspdf}
\usepackage{graphics}
\doi{10.1214/09-AOS761}
\volume{38}
\issue{3}
\pubyear{2010}
\firstpage{1593}
\lastpage{1606}

\makeatletter

\newtheorem{theorem}{Theorem}
\newtheorem{corollary}{Corollary}
\newproclaim{Algorithm}{Algorithm}
\newproclaim{Problem}{Problem}

\makeatother

\begin{document}
\begin{frontmatter}

\title{Monotonic convergence of a general algorithm for computing
optimal designs}
\runtitle{Algorithm for optimal designs}

\begin{aug}
\author[A]{\fnms{Yaming} \snm{Yu}\corref{}\ead[label=e1]{yamingy@uci.edu}}
\runauthor{Y. Yu}
\affiliation{University of California, Irvine}
\address[A]{Department of Statistics\\
University of California\\
Irvine, California, 92697-1250\\
USA\\
\printead{e1}} 
\end{aug}

\received{\smonth{5} \syear{2009}}
\revised{\smonth{10} \syear{2009}}

%
\begin{abstract}
Monotonic convergence is established for a general class of
multiplicative algorithms
introduced by Silvey, Titterington and Torsney
[\textit{Comm. Statist. Theory Methods} \textbf{14} (1978) 1379--1389] for
computing optimal designs. A conjecture of Titterington [\textit{Appl. Stat.}
\textbf{27} (1978) 227--234] is confirmed as a consequence. Optimal
designs for logistic regression are used as an illustration.
\end{abstract}

%

\end{frontmatter}

\section{A general class of algorithms}\label{sec1}
Optimal experimental design (approximate theory) is a well-developed
area, and we refer to Kiefer (\citeyear{Kiefer74}), Silvey (\citeyear
{Silvey80}), P\'{a}zman (\citeyear{Paz86})
and Pukelsheim (\citeyear{Pu93}) for a general introduction and basic
results. We
consider computational aspects of optimal designs, focusing on a finite
design space $\mathcal{X}=\{x_1, \ldots, x_n\}$. Suppose the
probability density or mass function of the response is specified as
$p(y|x, \theta)$ where $\theta=(\theta_1,\ldots, \theta_m)^\top$ is the
parameter of interest. Let $A_i$ denote the $m\times m$ expected Fisher
information matrix from a unit assigned to $x_i$ with the $(j,k)$ entry
[the expectation is with respect to $p(y|x_i, \theta)$]
\[
A_i(j,k)=E \biggl[\frac{\partial\log p(y|x_i, \theta)}{\partial\theta
_j}\, \frac{\partial\log p(y|x_i, \theta)}{\partial
\theta_k} \biggr].
\]
The moment matrix, as a function of the design measure $w=(w_1,\ldots,
w_n)$, is defined as
\[
M(w)=\sum_{i=1}^n w_i A_i
\]
which is proportional to the Fisher information for $\theta$ when the
number of units assigned to $x_i$ is proportional to $w_i$. Here $w\in
\bar{\Omega}$, and $\bar{\Omega}$ denotes the closure of $\Omega= \{
w\dvtx w_i> 0, \sum_{i=1}^n w_i=1 \}$.
Throughout we assume that $A_i$ are well defined and hence nonnegative
definite. The set
\[
\Omega_+\equiv\{w\in\bar{\Omega}\dvtx M(w)>0 \mbox{ (positive
definite)}\},
\]
is assumed nonempty. Our approach may conceivably extend to the case
where $M(w)$ is allowed to be singular, by using generalized inverses,
although we do not pursue this here.

Given an optimality criterion $\phi$ defined on positive definite
matrices, the goal is to maximize $\phi(M(w))$ with respect to $w\in
\Omega_+$. Typical optimality criteria include:
\begin{longlist}
\item the D-criterion $\phi_0(M)=\log\det(M)$,
\item the A-criterion $\phi_{-1}(M)=-\operatorname{tr}(M^{-1})$,
\item more generally, the $p$th mean criterion $\phi_p(M)=-\operatorname{tr}(M^p),
p<0$ and
\item the c-criterion $\phi_{-1,c}(M)=-c^\top M^{-1} c$, where $c$ is a
nonzero constant vector.
\end{longlist}

Often only a linear combination $K^\top\theta$, for example, a
subvector of $\theta$, is of interest. The Fisher information for
$K^\top\theta$ is naturally defined as $(K^\top M^{-1} K)^{-1}$,
assuming invertibility [Pukelsheim (\citeyear{Pu93})]. We may therefore consider
the D- and A-criteria for $K^\top\theta$ defined, respectively, as
%
\begin{eqnarray}\label{phik}
\phi_{0, K}(M) &=& -\log\det(K^\top M^{-1} K);\nonumber\\[-8pt]\\[-8pt]
\phi_{-1, K}(M) &=& -\operatorname{tr}(K^\top M^{-1} K).\nonumber
\end{eqnarray}
The c-criterion is a special case of $\phi_{-1, K}(M)$. Motivations for
such optimality criteria are well known. In a linear problem, the
A-criterion seeks to minimize the sum of variances of the best linear
unbiased estimators (BLUEs) for all coordinates of $\theta$ while the
c-criterion seeks to minimize the variance of the BLUE for $c^\top
\theta$. Similar interpretations (with asymptotic arguments) apply to
nonlinear problems.

In general $M(w)$ also depends on the unknown parameter $\theta$ which
complicates the definition of an optimality criterion. A simple
solution is to maximize $\phi(M(w))$ with $\theta$ fixed at a prior
guess $\theta^*$; this leads to \textit{local optimality} [Chernoff
(\citeyear{Chernoff53})]. Local optimality may be criticized for
ignoring uncertainty in
$\theta$. However, in a situation where real prior information is
available, or where the dependence of $M$ on $\theta$ is weak, it is
nevertheless a viable approach and has been adopted routinely [see,
e.g., Li and Majumdar (\citeyear{LM08})]. Henceforth we assume a fixed
$\theta^*$
and suppress the dependence of $M$ on $\theta$. Possible extensions are
mentioned in Section~\ref{sec5}.

Optimal designs do not usually come in closed form. As early as Wynn
(\citeyear{Wynn72}), Fedorov (\citeyear{Fed72}), Atwood (\citeyear{Atwood73})
and Wu and Wynn (\citeyear{WW78}), and as
late as Torsney (\citeyear{Torsney07}), Harman and Pronzato (\citeyear
{HP07}) and
Dette, Pepelyshev and Zhigljavsky (\citeyear{DPZ08}), various
procedures have been studied for numerical computation.
We shall focus on the following multiplicative algorithm [Titterington
(\citeyear{Titterington76}, \citeyear{Titterington78}),
Silvey, Titterington and Torsney (\citeyear{STT78})] which is specified
through a power
parameter $\lambda\in(0,1]$.
\renewcommand{\theAlgorithm}{I}
\begin{Algorithm}\label{AlgorithmI}
Set $\lambda\in(0, 1]$ and $w^{(0)}\in\Omega$. For $t=0,1,\ldots,$ compute
%
%
\begin{equation}
\label{alg1}
w_i^{(t+1)}= w_i^{(t)} \frac{d_i^\lambda(w^{(t)} )}{\sum
_{j=1}^n w_j^{(t)} d_j^\lambda(w^{(t)} )},\qquad i=1,\ldots, n,
\end{equation}
where
\[
d_i(w)=\operatorname{tr}(\phi'(M(w))A_i),\qquad \phi'(M)\equiv\frac{\partial\phi
(M)}{\partial M}.
\]
Iterate until convergence.

For a heuristic explanation, observe that (\ref{alg1}) is equivalent to
%
%
\begin{equation}
\label{heur}
w_i^{(t+1)}\propto w_i^{(t)} \biggl(\frac{\partial\phi
(M(w))}{\partial w_i} \bigg|_{w=w^{(t)}} \biggr)^\lambda,\qquad
i=1,\ldots, n.
\end{equation}
The value of $\partial\phi(M(w))/\partial w_i$ indicates the amount of
gain in information, as measured by $\phi$, by a slight increase in
$w_i$, the weight on the $i$th design point. So (\ref{heur}) can be
seen as adjusting $w$ so that relatively more weight is placed on
design points whose increased weight may result in a larger gain in
$\phi$. If $\phi$ is increasing and concave, then a convenient
convergence criterion, based on the general equivalence theorem
[Kiefer and Wolfowitz (\citeyear{KW60}), Whittle (\citeyear
{Whittle73})], is
%
%
\begin{equation}
\label{conv}
\max_{1\leq i\leq n} d_i \bigl(w^{(t)} \bigr) \leq(1+\delta) \bar
{d} \bigl(w^{(t)} \bigr),
\end{equation}
where $\bar{d}(w)\equiv\sum_{i=1}^n w_i d_i(w)$ and $\delta$ is a
small positive constant.

Algorithm I is remarkable in its generality. For example, little restriction is
placed on the underlying model $p(y|x, \theta)$. Part of the reason, of
course, is that we focus on Fisher information and local optimality,
which essentially reduces the problem to a linear one.

There exists a large literature on Algorithm \ref{AlgorithmI} and its relatives [see,
e.g., Titterington (\citeyear{Titterington76}, \citeyear{Titterington78}),
Silvey, Titterington and Torsney (\citeyear{STT78}), P\'{a}zman
(\citeyear{Paz86}), Fellman (\citeyear{Fellman89}), Pukelsheim and
Torsney (\citeyear{PT91}), Torsney and
Mandal (\citeyear{TM06}), Harman and Pronzato (\citeyear{HP07}), Dette,
Pepelyshev and Zhigljavsky (\citeyear{DPZ08}) and
Torsney and Mart\'{i}n-Mart\'{i}n (\citeyear{TM09})]. One feature that has
attracted much attention is that Algorithm \ref{AlgorithmI} appears to be monotonic,
that is, $\phi(M(w^{(t)}))$ increases in $t$, at least in some special
cases. For example, when $\phi=\phi_0$ (for D-optimality) and $\lambda
=1$, Titterington (\citeyear{Titterington76}) and P\'{a}zman (\citeyear
{Paz86}) have shown monotonicity
using clever probabilistic and analytic inequalities [see also Dette,
Pepelyshev and Zhigljavsky (\citeyear{DPZ08})
and Harman and Trnovsk\'{a} (\citeyear{HT09})]. Algorithm \ref{AlgorithmI} is also
known to be monotonic for $\phi=\phi_{-1, K}$ as in (\ref{phik}),
assuming $\lambda=1/2$ and $A_i$ are rank-one [Fellman (\citeyear
{Fellman74}), Torsney
(\citeyear{Torsney83})]. Monotonicity is important because convergence
then holds under
mild assumptions (see Section \ref{sec4}). Results in these special cases
suggest a monotonic convergence theory for a broad class of $\phi$
which is also supported by numerical evidence presented in some of the
references above.
\end{Algorithm}

\section{Main result}\label{sec2}
We aim to state general conditions on $\phi$ that ensure that Algorithm
\ref{AlgorithmI} converges monotonically. As a consequence certain known theoretical
results are unified and generalized, and one particular conjecture
[Titterington (\citeyear{Titterington78})] is confirmed. Define
\[
\psi(M)\equiv-\phi(M^{-1}),\qquad M>0.
\]
The functions $\phi$ and $\psi$ are assumed to be differentiable on
invertible matrices. Our conditions are conveniently stated in terms of
$\psi$. As usual, for two symmetric matrices, $M_1\leq(<)M_2$ means
$M_2-M_1$ is nonnegative (positive) definite.

\begin{itemize}
\item
$\psi(M)$ is increasing:
%
%
\begin{equation}
\label{INCREASE}
0< M_1\leq M_2 \quad\Longrightarrow\quad\psi(M_1)\leq\psi(M_2)
\end{equation}
or, equivalently, $\psi'(M)$ is nonnegative definite for positive
definite $M$.
\item
$\psi(M)$ is concave:
%
%
\begin{equation}
\label{CONCAVE0}
\alpha\psi(M_1)+(1-\alpha) \psi(M_2)\leq\psi\bigl(\alpha M_1+(1-\alpha) M_2\bigr)
\end{equation}
for $\alpha\in[0,1], M_1, M_2>0$. Equivalently,
%
%
\begin{equation}
\label{concave}
\psi(M_2)\leq\psi(M_1)+\operatorname{tr}\bigl(\psi'(M_1) (M_2-M_1)\bigr),\qquad M_1, M_2>0.
\end{equation}
\end{itemize}
Condition (\ref{INCREASE}) is usually satisfied by any reasonable
information criterion\break [Pukelsheim (\citeyear{Pu93})]. Also note that,
if (\ref
{INCREASE}) fails, then $\partial\phi(M(w))/\partial w_i$ on the
right-hand side of (\ref{heur}) is not even guaranteed to be
nonnegative. The real restriction is the concavity condition (\ref
{CONCAVE0}). For example, (\ref{CONCAVE0}) is not satisfied by $\psi
_p(M)=-\phi_p(M^{-1})$ (the $p$th mean criterion) when $p<-1$. [It is
usually assumed that $\phi(M)$, rather than $\psi(M)$, is concave.]
Nevertheless, (\ref{CONCAVE0}) \textit{is} satisfied by a wide range of
criteria, including the commonly used D-, A- or c-criteria [see cases
(i) and (ii) in the illustration of the main result below].

Our main result is as follows.
\begin{theorem}[(General monotonicity)]
\label{main}
Assume (\ref{INCREASE}) and (\ref{CONCAVE0}). Assume that in iteration
(\ref{alg1}), with $0<\lambda\leq1$, we have
\[
M\bigl(w^{(t)}\bigr)>0,\qquad \phi'\bigl(M\bigl(w^{(t)}\bigr)\bigr) \neq0\quad \mbox{and}\quad M\bigl(w^{(t+1)}\bigr)>0.
\]
Then
\[
\phi\bigl(M\bigl(w^{(t+1)}\bigr)\bigr)\geq\phi\bigl(M\bigl(w^{(t)}\bigr)\bigr).
\]
\end{theorem}

In other words, under mild conditions which ensure that (\ref{alg1}) is
well defined [specifically, the denominator in (\ref{alg1}) is
nonzero], (\ref{INCREASE}) and (\ref{CONCAVE0}) imply that (\ref{alg1})
never decreases the criterion $\phi$. Let us illustrate Theorem \ref
{main} with some examples. For simplicity, in (i)--(iv) we display
formulae for $\lambda=1$ only, although monotonicity holds for all
$\lambda\in(0, 1]$.

\begin{longlist}
\item Take
\[
\phi_{p}(M)=
\cases{
\log\det M, &\quad $p=0$;\cr
-\operatorname{tr}(M^p), &\quad $p\in[-1, 0)$.}
\]
Then $\psi_p(M)\equiv-\phi_p(M^{-1})$ satisfies (\ref{INCREASE}) and
(\ref{CONCAVE0}). By Theorem \ref{main}, Algorithm \ref{AlgorithmI} is monotonic for
$\phi=\phi_p, p\in[-1, 0]$. This generalizes the previously known
cases $p=0$ and $p=-1$ (with particular values of $\lambda$). The
iteration (\ref{alg1}) reads
\[
w_i^{(t+1)}= w_i^{(t)} \frac{\operatorname{tr}(M^{p-1}(w^{(t)})
A_i)}{\operatorname{tr}(M^{p}(w^{(t)}))},\qquad i=1, \ldots, n.
\]

\item
More generally, given a full rank $m\times r$ matrix $K$ ($r\leq m$), consider
\[
\psi_{p, K}(M^{-1})\equiv-\phi_{p, K}(M)=
\cases{
\log\det(K^\top M^{-1} K), &\quad $p=0$;\cr
\operatorname{tr}((K^\top M^{-1} K)^{-p}), &\quad $p\in[-1, 0)$.}
\]
Then $\psi_{p, K}(M)$ satisfies (\ref{INCREASE}) and (\ref{CONCAVE0}).
By Theorem \ref{main}, Algorithm \ref{AlgorithmI} is monotonic for $\phi=\phi_{p, K},
p\in[-1, 0]$. The iteration (\ref{alg1}) reads
%
%
\begin{equation}
\label{pK}
w_i^{(t+1)}= w_i^{(t)} \frac{\operatorname{tr}(M^{-1}K(K^\top M^{-1}K)^{-p-1}
K^\top M^{-1} A_i)}{\operatorname{tr}((K^\top M^{-1}K)^{-p} )} \bigg|_{M=M(w^{(t)})}.
\end{equation}

\item  In particular, taking $r=1, K=c$ (an $m\times1$ vector) and
$p=-1$ in case~(ii), we obtain that Algorithm \ref{AlgorithmI} is monotonic for the
c-criterion $\phi_{-1,c}$. The iteration (\ref{pK}) reduces to
\[
w_i^{(t+1)}= w_i^{(t)} \frac{c^\top M^{-1}(w^{(t)}) A_i
M^{-1}(w^{(t)})c}{c^\top M^{-1}(w^{(t)})c},\qquad i=1, \ldots, n.
\]

As noted by a referee, with $p=-1$, the choice $\lambda=1$ may lead to
an oscillating behavior in the sense that $w^{(t)}$ alternates between
two points at which $\phi_{-1,c}(M(w))$ takes the same value. While
this does not contradict Theorem \ref{main}, it suggests that other
values of $\lambda$ are more desirable for fast convergence. Following
Fellman (\citeyear{Fellman74}) and Torsney (\citeyear{Torsney83}),
a practical recommendation is $\lambda=1/2$ in the $p=-1$ case.

\item  Consider another example of case (ii), with $p=0, r=m-1$ and
$K=(0_r, I_r)^\top$. Henceforth $0_r$ denotes the $r\times1$ vector of
zeros, and $I_r$ denotes the $r\times r$ identity matrix. Assume
$A_i=x_ix_i^\top, x_i^\top=(1, z_i^\top)$ and $z_i$ is $(m-1)\times
1$. This corresponds to a D-optimal design problem for $(\theta_2,
\ldots, \theta_{m})$ under the linear model,
\[
y|(x, \theta)\sim\mathrm{N}(x^\top\theta, \sigma^2),\qquad x^\top=(1,
z^\top),
\]
where the parameter is $\theta=(\theta_1, \theta_2,\ldots, \theta
_{m})^\top$. That is, interest centers on all coefficients other than
the intercept. Nevertheless, as far as the design measure $w$ is
concerned, the optimality criterion, $\phi_{0, K}(M)$, coincides with
$\phi_0(M)$, that is,
\[
-\log\det(K^\top M^{-1}(w) K)=\log\det M(w).
\]
After some algebra, (\ref{pK}) reduces to
%
%
\begin{equation}
\label{alg2}
w_i^{(t+1)}=w_i^{(t)} \frac{(z_i-\bar{z})^{\top} M_c^{-1}(w^{(t)})
(z_i-\bar{z})}{m-1},\qquad i=1, \ldots, n,
\end{equation}
where
\[
\bar{z}=\sum_{i=1}^n w_i^{(t)} z_i;\qquad M_c\bigl(w^{(t)}\bigr)=\sum_{i=1}^n
w_i^{(t)} (z_i-\bar{z}) (z_i-\bar{z})^\top.
\]
Thus (\ref{alg2}) satisfies $\det M(w^{(t+1)})\geq\det M(w^{(t)})$.
\end{longlist}

Monotonicity of (\ref{alg2}) has been conjectured since Titterington
(\citeyear{Titterington78}), and considerable numerical evidence has
accumulated over the
years. Recently, extending the arguments of P\'{a}zman (\citeyear
{Paz86}), Dette, Pepelyshev
and Zhigljavsky (\citeyear{DPZ08}) have obtained results which come very
close to resolving
Titterington's conjecture. Nevertheless, we have been unable to extend
their arguments further. Instead we prove the general Theorem \ref
{main} using a different approach, and settle this conjecture as a consequence.

The proof of Theorem \ref{main} is achieved by using a method of
\textit{auxiliary variables}. When a function $f(w)$ [e.g., $-\det M(w)$] to be
minimized is complicated, we introduce a new variable $Q$ and a
function $g(w, Q)$ such that $\min_Q g(w, Q)=f(w)$ for all $w$, thus
transforming the problem into minimizing $g(w, Q)$ over $w$ and $Q$
jointly. Then we may use an iterative conditional minimization strategy
on $g(w, Q)$. This is inspired by the EM algorithm [Dempster, Laird and Rubin
(\citeyear{DLR77}), Meng and van Dyk (\citeyear{MV97}); in particular,
see Csisz\'{a}r and
Tusnady's (\citeyear{CT84}) interpretation; see Yu (\citeyear{Yu08})
for a related
interpretation of the data augmentation algorithm].

In Section \ref{sec3} we analyze Algorithm \ref{AlgorithmI} using this strategy. Although
attention is paid to the mathematics, our focus is on intuitively
appealing interpretations which may lead to further extensions of
Algorithm \ref{AlgorithmI} with the same desirable monotonicity properties. If the
algorithm is monotonic, then convergence can be established under mild
conditions (Section \ref{sec4}). Section \ref{sec5} contains an illustration with optimal
designs for a simple logistic regression model.

\section{Explaining the monotonicity}\label{sec3}
A key observation is that the problem of maximizing $\phi(M(w))$, or,
equivalently, minimizing $\psi(M^{-1}(w))$ can be formulated as a
joint minimization over both the design and the estimator.
Specifically, let us compare the original Problem \ref{ProblemP1} with its companion
Problem \ref{ProblemP2}. Throughout $A^{1/2}$ denotes the symmetric nonnegative definite
(SNND) square root of an SNND matrix $A$.
\renewcommand{\theProblem}{P\arabic{Problem}}
\begin{Problem}\label{ProblemP1}
Minimize $-\phi(M(w))\equiv\psi((\sum_{i=1}^n w_i
A_i)^{-1})$ over $w\in\Omega$.
\end{Problem}
\begin{Problem}\label{ProblemP2}
Minimize
%
%
\begin{equation}
\label{gwQ}
g(w, Q)\equiv\psi(Q \Delta_{w}Q^\top)
\end{equation}
over $w\in\Omega$ and $Q$ [an $m\times(mn)$ matrix], subject to
$QG=I_{m}$, where
\[
\Delta_w \equiv\operatorname{Diag}(w^{-1}_1, \ldots, w^{-1}_n)\otimes
I_m;\qquad
G \equiv(A_1^{1/2}, \ldots, A_n^{1/2})^\top.
\]
\end{Problem}

Though not immediately obvious, Problems \ref{ProblemP1} and \ref{ProblemP2} are equivalent, and this may
be explained in statistical terms as follows. In (\ref{gwQ}), $Q\Delta
_w Q^\top$ is simply the variance matrix of a linear unbiased
estimator, $QY$, of the $m\times1$ parameter $\theta$ in the model
\[
Y=G\theta+\varepsilon,\qquad \varepsilon\sim\mathrm{N}(0, \Delta_w),
\]
where $Y$ is the $(mn)\times1$ vector of observations. The constraint
$QG=I_{m}$ ensures unbiasedness. [Note that $G$ is full-rank since
$M(w)$ is nonsingular by assumption.] Of course, the weighted least
squares (WLS) estimator is the best linear unbiased estimator, having
the smallest variance matrix (in the sense of positive definite
ordering) and, by (\ref{INCREASE}), the smallest $\psi$ for that
matrix. It follows that, for fixed $w$, $g(w, Q)$ is minimized by
choosing $QY$ as the WLS estimator,
%
\begin{eqnarray}
\label{gmin1}
g(w, \hat{Q}_{\mathrm{WLS}}) &=& \inf_{QG=I_{m}} g(w, Q),\\
\label{Qwls}
\hat{Q}_{\mathrm{WLS}} &=& M^{-1}(w) (w_1 A_1^{1/2},\ldots, w_n A_n^{1/2} ).
\end{eqnarray}
However, from (\ref{gwQ}) and (\ref{Qwls}) we get
%
%
\begin{equation}
\label{gmin2}
g(w, \hat{Q}_{\mathrm{WLS}})=\psi(M^{-1}(w)).
\end{equation}
That is, Problem \ref{ProblemP2} reduces to Problem \ref{ProblemP1} upon minimizing over $Q$.

Since Problem \ref{ProblemP2} is not immediately solvable, it is natural to consider the
subproblems: (i) minimizing $g(w, Q)$ over $Q$ for fixed $w$ and (ii)
minimizing $g(w, Q)$ over $w$ for fixed $Q$. Part (ii) is again
formulated as a joint minimization problem. For a fixed $m\times(mn)$
matrix $Q$ such that $QG=I_m$, let us consider Problems \ref{ProblemP3} and \ref{ProblemP4}.
\begin{Problem}\label{ProblemP3}
Minimize $g(w, Q)$ as in (\ref{gwQ}) over $w\in\Omega$.
\end{Problem}
\begin{Problem}\label{ProblemP4}
Minimize the function
%
%
\begin{equation}
\label{hmin2}
h(\Sigma, w, Q)=\psi(\Sigma) +\operatorname{tr}\bigl(\psi'(\Sigma) (Q \Delta_w
Q^\top-\Sigma) \bigr),
\end{equation}
over $w\in\Omega$ and the $m\times m$ positive-definite matrix $\Sigma$.
\end{Problem}

The concavity assumption (\ref{concave}) implies that
%
%
\begin{equation}
\label{hmin1}
h(\Sigma, w, Q)\geq\psi(Q \Delta_w Q^\top)
\end{equation}
with equality when $\Sigma=Q \Delta_w Q^\top$, that is, Problem \ref{ProblemP4}
reduces to Problem \ref{ProblemP3} upon minimizing over $\Sigma$.

Since Problem \ref{ProblemP4} is not immediately solvable, it is natural to consider the
subproblems: (i) minimizing $h(\Sigma, w, Q)$ over $\Sigma$ for fixed
$w$ and $Q$ and (ii) minimizing $h(\Sigma, w, Q)$ over $w$ for fixed
$\Sigma$ and $Q$. Part (ii), which amounts to minimizing
\[
\operatorname{tr}(\psi'(\Sigma) Q\Delta_w Q^\top)=\operatorname{tr}(Q^\top\psi
'(\Sigma) Q \Delta_w ),
\]
admits a closed-form solution: if we write $Q=(Q_1, \ldots, Q_n)$ where
each $Q_i$ is $m\times m$, then
$w_i^2$ should be proportional to $\operatorname{tr}(Q_i^\top\psi'(\Sigma) Q_i)$. But
Algorithm \ref{AlgorithmI} may not perform an exact minimization here [see (\ref{s3})].

Based on the above discussion, we can express Algorithm \ref{AlgorithmI} as an
iterative conditional minimization algorithm involving $w, Q$ and
$\Sigma$. At iteration $t$, define
\begin{eqnarray*}
Q^{(t)} &=& \bigl(Q_1^{(t)}, \ldots, Q_n^{(t)}\bigr);\\
Q_i^{(t)} &=& w_i^{(t)} M^{-1}\bigl(w^{(t)}\bigr) A_i^{1/2},\qquad i=1,\ldots, n;\\
\Sigma^{(t)} &=& Q^{(t)} \Delta_{w^{(t)}} Q^{(t)\top} = M^{-1}\bigl(w^{(t)}\bigr).
\end{eqnarray*}
Then we have
%
\begin{eqnarray}
\psi\bigl(M^{-1}\bigl(w^{(t)}\bigr)\bigr) &=& g\bigl(w^{(t)}, Q^{(t)}\bigr)
\qquad\mbox{[by (\ref{gmin2})]}\nonumber\\
&=& h\bigl(\Sigma^{(t)}, w^{(t)}, Q^{(t)}\bigr)
\qquad\mbox{[by (\ref{hmin2})]}\nonumber\\
\label{s3}
&\geq& h\bigl(\Sigma^{(t)}, w^{(t+1)}, Q^{(t)}\bigr)
\qquad\mbox{(see below)}\\
\label{s4}
&\geq& g\bigl(w^{(t+1)}, Q^{(t)}\bigr)
\qquad\mbox{[by (\ref{hmin1}), (\ref{gwQ})]}\\
\label{s5}
&\geq& \psi\bigl(M^{-1}\bigl(w^{(t+1)}\bigr)\bigr)
\qquad\mbox{[by (\ref{gmin1}), (\ref{gmin2})]}.
\end{eqnarray}
The choice of $w^{(t+1)}$ leads to (\ref{s3}) as follows. After simple
algebra, the iteration (\ref{alg1}) becomes
\[
w^{(t+1)}_i = \frac{r_i^\lambda w_i^{1-2\lambda}}{\sum_{j=1}^n
r_j^\lambda w_j^{1-2\lambda}},\qquad i=1,\ldots, n,
\]
where
\[
w_i\equiv w_i^{(t)},\qquad r_i\equiv \operatorname{tr}\bigl(Q_i^{(t)\top}\psi'\bigl(\Sigma
^{(t)}\bigr)Q_i^{(t)} \bigr).
\]
Since $0< \lambda\leq1$, Jensen's inequality yields
\[
\Biggl(\sum_{i=1}^n \frac{r_i}{w_i} \Biggr)^{1-\lambda}\geq\sum_{i=1}^n
w_i \biggl(\frac{r_i}{w_i^2} \biggr)^{1-\lambda};\qquad \Biggl(\sum_{i=1}^n
\frac{r_i}{w_i} \Biggr)^\lambda\geq\sum_{i=1}^n w_i \biggl(\frac
{r_i}{w_i^2} \biggr)^\lambda.
\]
That is,
\[
\sum_{i=1}^n \frac{r_i}{w_i}\geq\Biggl(\sum_{i=1}^n r_i^{1-\lambda}
w_i^{2\lambda-1} \Biggr) \Biggl(\sum_{i=1}^n r_i^\lambda w_i^{1-2\lambda
} \Biggr).
\]
Hence
\begin{eqnarray*}
\operatorname{tr}\bigl(\psi'\bigl(\Sigma^{(t)}\bigr) Q^{(t)} \Delta_{w^{(t)}} Q^{(t)\top}
\bigr) &=& \sum_{i=1}^n \frac{r_i}{w_i^{(t)}}
\geq \sum_{i=1}^n \frac{r_i}{w_i^{(t+1)}}\\
&=&\operatorname{tr}\bigl(\psi'\bigl(\Sigma^{(t)}\bigr)
Q^{(t)} \Delta_{w^{(t+1)}} Q^{(t)\top} \bigr),
\end{eqnarray*}
which produces (\ref{s3}). Choosing $\lambda=1/2$, that is,
$w_i^{(t+1)}\propto\sqrt{r_i}$, leads to exact minimization in (\ref
{s3}); choosing $\lambda=1$ yields equality in (\ref{s3}). But any
choice of $w^{(t+1)}$ that decreases $h(\Sigma^{(t)}, w, Q^{(t)})$ at
(\ref{s3}) would have resulted in the desired inequality,
\[
\psi\bigl(M^{-1}\bigl(w^{(t)}\bigr)\bigr)\geq\psi\bigl(M^{-1}\bigl(w^{(t+1)}\bigr)\bigr).
\]
We may allow $\lambda$ to change from iteration to iteration, and
monotonicity still holds, as long as $\lambda\in(0, 1]$. See Silvey,
Titterington and Torsney (\citeyear{STT78}) and Fellman (\citeyear{Fellman89})
for investigations concerning the choice
of $\lambda$. Also note that we assume $w_i^{(t)}, w_i^{(t+1)}>0$ for
all $i$. This is not essential, however, because (i) the possibility of
$w_i^{(t)}=0$ can be handled by restricting our analysis to all design
points $i$ such that $w_i^{(t)}>0$, and (ii) the possibility of
$w_i^{(t+1)}=0$ can be handled by a standard limiting argument.
Monotonicity holds as long as $M(w^{(t)})$ and $M(w^{(t+1)})$ are both
positive definite, as noted in the statement of Theorem \ref{main}.

\section{Global convergence}\label{sec4}
Monotonicity (Theorem \ref{main}) plays an important role in the
following convergence theorem.
\begin{theorem}[(Global convergence)]
\label{THM2}
Denote the mapping (\ref{alg1}) by $T$.

\textup{(a)} Assume
\[
\phi'(M(w))\geq0;\qquad \phi'(M(w))A_i \neq0,\qquad w\in\Omega_+,
i=1,\ldots, n.
\]

\textup{(b)} Assume (\ref{alg1}) is strictly monotonic, that is,
%
%
\begin{equation}
\label{iff}
w\in\Omega_+,\qquad Tw\neq w \quad\Longrightarrow\quad\phi(M(Tw))>\phi(M(w)).
\end{equation}

\textup{(c)} Assume $\phi$ is strictly concave and $\phi'$ is continuous on
positive definite matrices.

\textup{(d)} Assume that, if $M$ (a positive definite matrix) tends to $M^*$
such that $\phi(M)$ increases monotonically, then $M^*$ is nonsingular.

Let $w^{(t)}$ be generated by (\ref{alg1}) with $w^{(0)}_i>0$ for all
$i$. Then:

\begin{longlist}
\item all limit points of $w^{(t)}$ are global maxima of $\phi(M(w))$ on
$\Omega_+$, and

\item as $t\to\infty$, $\phi(M(w^{(t)}))$ increases monotonically to
$\sup_{w\in\Omega_+} \phi(M(w))$.
\end{longlist}
\end{theorem}

The proof of Theorem \ref{THM2} is somewhat subtle. Standard arguments
show that all limit points of $w^{(t)}$ are fixed points of the mapping
$T$. This alone does not imply convergence to a global maximum,
however, because there often exist sub-optimal fixed points on the
boundary of $\Omega$. (Global maxima occur routinely on the boundary
also.) Our goal is therefore to rule out possible convergence to such
sub-optimal points; details of the proof are presented in Yu (\citeyear
{Yu09}), an
extended version of this paper. We shall comment on conditions (a)--(d).

Condition (a) ensures that starting with $w^{(0)}\in\Omega_+$, all
iterations are well defined. Moreover, if $w^{(0)}_i>0$ for all $i$,
then $w^{(t)}_i>0$ for all $t$ and $i$. This highlights the basic idea
that, in order to converge to a global maximum $w^*$, the starting
value $w^{(0)}$ must assign positive weight to every support point of
$w^*$. Such a requirement is not necessary for monotonicity. On the
other hand, assigning weight to nonsupporting points of $w^*$ tends to
slow the algorithm down. Hence methods that quickly eliminate
nonoptimal support points are valuable [Harman and Pronzato (\citeyear{HP07})].

Condition (b) simply says that unless $w$ is a fixed point, the mapping
$T$ should produce a better solution.
Let us assume (\ref{INCREASE}), (\ref{concave}) and condition (a) so
that Theorem~\ref{main} applies. Then, by checking the equality
condition in (\ref{s3}), it is easy to see that condition (b) is
satisfied if $0<\lambda<1$. [The argument leading to (\ref{iff})
technically assumes that all coordinates of $w$ are nonzero, but we can
apply it to the appropriate subvector of $w$.] If $\lambda=1$, then
(\ref{s3}) reduces to an equality. However, by checking the equality
conditions in (\ref{s4}) and (\ref{s5}), we can show that condition (b)
is satisfied if
$\psi$ is strictly increasing and strictly concave:
%
\begin{eqnarray}\qquad
\label{sincrease}
&&M_2 \geq M_1>0,\nonumber\\[-8pt]\\[-8pt]
&&\qquad M_1\neq M_2 \quad \Longrightarrow\quad \psi(M_1)<\psi
(M_2);\nonumber\\
\label{sconcave}
&&M_1, M_2>0,\nonumber\\[-8pt]\\[-8pt]
&&\qquad M_1\neq M_2 \quad\Longrightarrow\quad\psi(M_2)<\psi
(M_1)+\operatorname{tr}\bigl(\psi'(M_1)(M_2-M_1)\bigr).\nonumber
\end{eqnarray}

Conditions (c) and (d) are technical requirements that concern $\phi$
alone. Condition (c) ensures uniqueness of the optimal moment matrix
which simplifies the analysis. Condition (d) ensures that positive
definiteness of $M(w)$ is maintained in the limit. Conditions (c) and
(d) are satisfied by $\phi=\phi_p$ with $p\leq0$, for example.

Let us mention a typical example of Theorem \ref{THM2}.
\begin{corollary}
\label{coro1}
Assume $A_i\neq0, w^{(0)}_i>0, i=1,\ldots,n$, and $M(w^{(0)})>0$.
Then the conclusion of Theorem \ref{THM2} holds for Algorithm \ref{AlgorithmI} with
$\phi=\phi_0$.
\end{corollary}
\begin{pf}
Conditions (a), (c) and (d) are readily verified. Condition (b) is
satisfied by (\ref{sincrease}) and (\ref{sconcave}).
The claim follows from Theorem \ref{THM2}.
\end{pf}

When (\ref{sincrease}) or (\ref{sconcave}) fails, and $\lambda=1$, it
is often difficult to appeal to Theorem \ref{THM2} because strict
monotonicity [condition (b)] may not hold. We illustrate this with an
example where the monotonicity is not strict, and the algorithm does
not converge [see Pronzato, Wynn and Zhigljavsky (\citeyear{PrWyZh2000}), Chapter 7; also the remark in case (iii) following Theorem \ref
{main}]. Consider iteration (\ref{alg2}) \mbox{($\lambda=1$)} with $n=m=2$ and
design space $\mathcal{X}=\{x_i=(1, z_i)^\top, i=1,2\}, z_1=-z_2=1$.
It is easy to show that, for any $w^{(t)}=(w_1, w_2)\in\Omega$,
iteration (\ref{alg2}) maps $w^{(t)}$ to $w^{(t+1)}=(w_2, w_1)$. Thus,
unless $w_1=w_2=1/2$ to begin with, the algorithm alternates between
two distinct points. This appears to be a rare example, as (\ref{alg2})
usually converges in practical situations.


\section{Further remarks and illustrations}\label{sec5}
One can think of several reasons for the wide interest in Algorithm \ref{AlgorithmI}
and its relatives. Similar to the EM algorithm, Algorithm \ref{AlgorithmI} is simple,
easy to implement and monotonically convergent for a large class of
optimality criteria (although this was not proved in the present
generality). Algorithm \ref{AlgorithmI} is known to be slow sometimes. But it serves
as a foundation upon which more effective variants can be built [see,
e.g., Harman and Pronzato (\citeyear{HP07}) and Dette, Pepelyshev and
Zhigljavsky (\citeyear{DPZ08})]. While
solving the conjectured monotonicity of (\ref{alg2}) holds mathematical
interest, our main contribution is a way of interpreting such
algorithms as optimization on augmented spaces. This opens up new
possibilities in constructing algorithms with the same desirable
monotonic convergence properties.

As a numerical example, consider the logistic regression model
\[
p(y|x, \theta)=\frac{\exp(yx^\top\theta)}{1+\exp(x^\top\theta)},\qquad
y=0, 1.
\]
The expected Fisher information for $\theta$ from a unit assigned to
$x_i$ is
\[
A_i=x_i \frac{\exp(x_i^\top\theta)}{(1+\exp(x_i^\top\theta))^2}x_i^\top.
\]
We compute locally optimal designs with prior guess $\theta^*=(1,
1)^\top(m=2)$, and design spaces,
\begin{eqnarray*}
\mathcal{X}_1 &=& \{x_i=(1, i/20)^\top, i=1,\ldots, 20\};\\
\mathcal{X}_2 &=& \{x_i=(1, i/10)^\top, i=1,\ldots, 30\}.
\end{eqnarray*}
The design criteria considered are $\phi_0$ (for D-optimality) and $\phi
_{-2}$. We use Algorithm \ref{AlgorithmI} with $\lambda=1$, starting with equally
weighted designs.

\begin{figure}

\includegraphics{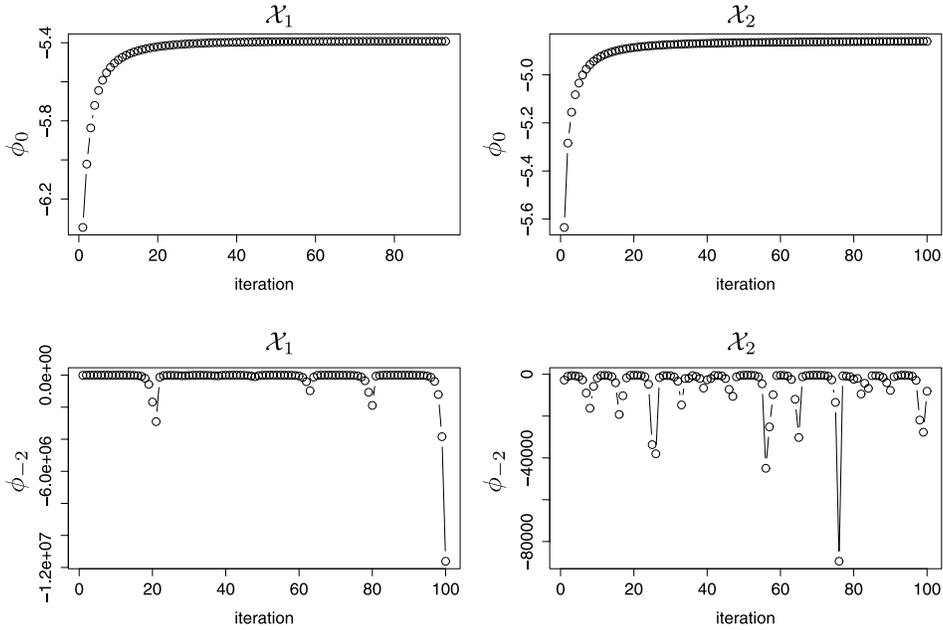}

\caption{Values of $\phi_0=\log\det M$ and
$\phi_{-2}=-\operatorname{tr}(M^{-2})$ for
Algorithm \protect\ref{AlgorithmI} with design spaces
$\mathcal{X}_1$ and $\mathcal{X}_2$.}
\label{fig1}
\end{figure}

For $\phi_0$, Corollary \ref{coro1} guarantees monotonic convergence.
This is illustrated by Figure \ref{fig1}, the first row, where $\phi
_0=\log\det M(w)$ is plotted against iteration $t$. Using the
convergence criterion (\ref{conv}) with $\delta=0.0001$, the number of
iterations until convergence is 93 for $\mathcal{X}_1$ and 2121 for
$\mathcal{X}_2$. The actual locally D-optimal designs are
$w_1=w_{20}=0.5$ for $\mathcal{X}_1$ and $w_1=w_{23}=0.5$ for $\mathcal
{X}_2$, as can be verified using the general equivalence theorem. This
simple example serves to illustrate both the monotonicity of Algorithm
\ref{AlgorithmI} (when Theorem \ref{main} applies) and its potential slow convergence.

%

For $\phi_{-2}$, although Algorithm \ref{AlgorithmI} can be implemented just as
easily, Theorem \ref{main} does not apply because the concavity
condition (\ref{concave}) no longer holds. Indeed, Algorithm \ref{AlgorithmI} (with
$\lambda=1$) is not monotonic, as is evident from Figure \ref{fig1}, in
the second row, where $\phi_{-2}=-\operatorname{tr}(M^{-2}(w))$ is plotted against
iteration $t$. This shows the potential danger of using Algorithm \ref{AlgorithmI}
when monotonicity is not guaranteed.

Although Theorem \ref{main} does not cover the $\phi_p$ criterion for
$p<-1$, it is still possible that
monotonicity holds for a smaller range of $\lambda$. Calculations in
special cases lead to the conjecture [Silvey, Titterington and Torsney
(\citeyear{STT78})] that
Algorithm \ref{AlgorithmI} is monotonic if $0<\lambda\leq1/(1-p)$. Theorem \ref{main}
provides further evidence for this conjecture, but new insights are
needed to resolve it.

We have focused on local optimality. An alternative, \textit{Bayesian
optimality} [Chaloner and Larntz (\citeyear{CL89}), Chaloner and Verdinelli
(\citeyear{CV95})], seeks to maximize the expected value of $\phi
(M(\theta; w))$
over a prior distribution $\pi(\theta)$. The notation $M(\theta; w)$
emphasizes the dependence of the moment matrix on the parameter $\theta
$. It would be worthwhile to extend our strategy in Section \ref{sec3} to
Bayesian optimality, and we plan to report both theoretical and
empirical evaluations of such extensions in future works.

\section*{Acknowledgment}
The author would like to thank  Don Rubin, Xiao-Li Meng
and David van Dyk for introducing him to the field of statistical
computing. He is also grateful to  Mike Titterington, Ben
Torsney and the referees for their valuable comments.

\printaddresses

\end{document}